\documentclass[
 aip,
 amsmath,
 amssymb,
 reprint
]{revtex4-1}

\usepackage{graphicx}
\usepackage{dcolumn}
\usepackage{bm}

\usepackage[utf8]{inputenc}
\usepackage[T1]{fontenc}
\usepackage{mathptmx}
\usepackage{etoolbox}
\usepackage{url}
\usepackage{multirow}
\usepackage{makecell}
\usepackage{float}

\makeatletter
\def\@email#1#2{%
 \endgroup
 \patchcmd{\titleblock@produce}
  {\frontmatter@RRAPformat}
  {\frontmatter@RRAPformat{\produce@RRAP{*#1\href{mailto:#2}{#2}}}\frontmatter@RRAPformat}
  {}{}
}%
\makeatother
\begin{document}

\preprint{AIP/123-QED}

\title{Magnetic characterization of electronic components for portable atomic sensors using a zero-field optically pumped magnetometry platform}

\author{Hyeonjae Kim}
\affiliation{3\textsuperscript{rd} R\&D Institute, Agency for Defense Development, Daejeon 34186, Republic of Korea}
\author{Sangkyung Lee}
\thanks{Corresponding author: sklee82@add.re.kr}
\affiliation{3\textsuperscript{rd} R\&D Institute, Agency for Defense Development, Daejeon 34186, Republic of Korea}
\author{Sin Hyuk Yim}
\affiliation{3\textsuperscript{rd} R\&D Institute, Agency for Defense Development, Daejeon 34186, Republic of Korea}
\author{Taek Jeong}
\affiliation{3\textsuperscript{rd} R\&D Institute, Agency for Defense Development, Daejeon 34186, Republic of Korea}
\author{Younghoon Lim}
\affiliation{3\textsuperscript{rd} R\&D Institute, Agency for Defense Development, Daejeon 34186, Republic of Korea}

\date{\today}

\begin{abstract}
We present a zero-field optically pumped magnetometry platform for magnetic characterization of a photodetector (PD) board and a resistance temperature detector (RTD) used in a portable atomic magnetometer. For each component, the static magnetic field along the measurement axis is determined from the shift in the center of the dispersive response, while response distortion caused by off-axis magnetic-field components is assessed from the absorptive admixture. Magnetic-field noise is evaluated based on the quadrature difference in the amplitude spectral density (ASD). The platform achieved a $-3$~dB bandwidth of 71.5~Hz and a median ASD of 57.6~fT/$\sqrt{\mathrm{Hz}}$ over 20--70~Hz. For the unpowered PD board, static magnetic fields of approximately 8~nT in magnitude and opposite signs were measured in the front- and back-facing orientations. The back-facing orientation also exhibited a degraded dispersive response consistent with magnetic-field inhomogeneity. The magnetic-noise contributions associated with the PD board were estimated at 36.0 and 63.2~fT/$\sqrt{\mathrm{Hz}}$ in the front- and back-facing orientations, respectively. By contrast, operation of the RTD readout generated a static magnetic field of $-0.6$~nT, with no measurable response degradation or additional magnetic-field noise. These results provide guidance for the design and placement of electronic components in portable atomic magnetometers targeting sensitivities below 0.1~pT/$\sqrt{\mathrm{Hz}}$.
\end{abstract}
\maketitle

\section{Introduction}

Optically pumped magnetometers (OPMs) are atomic sensors that measure magnetic fields through the spin-polarization response of an atomic ensemble.\cite{Budker2007optical} Providing high sensitivity without cryogenic cooling, OPMs have been used in diverse applications, including magnetoencephalography,\cite{Boto2018moving} magnetocardiography,\cite{Wyllie2012magnetocardiography} magnetic anomaly detection,\cite{Zhou2026research} and magnetic navigation.\cite{Canciani2017airborne} A major direction in OPM development has been to reduce sensor size while maintaining high sensitivity. Operation in the spin-exchange relaxation-free (SERF) regime, with high alkali-metal vapor density and near-zero magnetic field, enables femtotesla-level sensitivity.\cite{Allred2002high,Kominis2003subfemtotesla} In parallel, advances in microfabrication have led to chip-scale atomic vapor cells that can be integrated with miniaturized optical and electronic components into compact sensor packages.\cite{Liew2004microfabricated,Kitching2018chip} NIST demonstrated this combination in a micromachined SERF OPM with a $0.36~\mathrm{cm^3}$ sensor head and a sensitivity below $20~\mathrm{fT}/\sqrt{\mathrm{Hz}}$ over most of the 15--100~Hz range.\cite{Mhaskar2012low} More recently, researchers at Beihang University developed a $10~\mathrm{cm^3}$ single-beam SERF OPM with a sensitivity of $20~\mathrm{fT}/\sqrt{\mathrm{Hz}}$.\cite{Yin2022influence}

Miniaturization of sensor heads brings electronic components closer to the atomic vapor cell, thereby increasing the influence of component-induced magnetic perturbations on the atomic ensemble.\cite{Preusser2009microfabricated} Even without power, magnetic materials in electronic components can retain remanent magnetization and produce static magnetic fields. The component of such a static magnetic field along the measurement axis can introduce heading error by altering the total magnetic field relative to the spin-polarization axis, resulting in an angle-dependent error in the measured magnetic field.\cite{Lee2021heading} Spatial variations in static magnetic-field components, including those along the other axes, can induce magnetic-field inhomogeneity within the atomic vapor cell, thereby degrading the sensor response.\cite{Fang2022analysis} Moreover, thermal magnetization fluctuations in magnetic materials can generate magnetic noise,\cite{Lee2008calculation} raising the noise floor and limiting sensitivity.\cite{Karaulanov2016spin} When powered, operating currents and their fluctuations can generate additional magnetic fields and magnetic noise, respectively. Characterization of both static magnetic fields and magnetic-field noise is therefore essential for the design and integration of electronic components in compact, high-sensitivity OPMs.

Previous studies have used atomic magnetometers to characterize magnetic fields generated by electronic components. Lu \textit{et al.} measured the three-axis magnetic fields from an operating heater and temperature sensor using the required compensation fields.\cite{Lu2020insitu} Liang \textit{et al.} measured the three-axis magnetic fields from an integrated heating film using shifts in the noble-gas Larmor frequency.\cite{Liang2021magnetic} Hunter \textit{et al.} used free-induction-decay (FID) measurements with optical beam steering to image magnetic fields from operating electronic components, including printed circuit-board traces, a bridge rectifier, and a ceramic battery.\cite{Hunter2026high} However, these studies focused on component-generated magnetic fields without evaluating the additional magnetic-field noise associated with the tested components.

\begin{figure*}
    \centering
    \includegraphics[width=\textwidth]{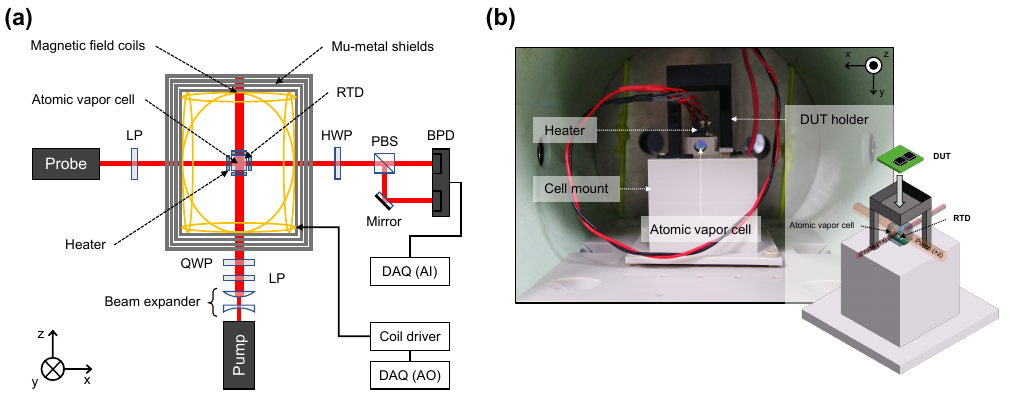}
    \caption{Configuration of the zero-field optically pumped magnetometry platform. (a) Schematic of the optical and electronic setup. The pump and probe beams propagate along the $z$ and $x$ axes, respectively. LP, linear polarizer; QWP, quarter-wave plate; HWP, half-wave plate; PBS, polarizing beam splitter; BPD, balanced photodetector; DAQ, data acquisition; AO, analog output; AI, analog input; RTD, resistance temperature detector. (b) Photograph of the atomic vapor cell and the device under test (DUT) holder inside the innermost mu-metal shield, together with an illustration of the DUT placement in the holder.}
    \label{fig:setup}
\end{figure*}

In this work, we developed a zero-field optically pumped magnetometry platform for magnetic characterization of electronic components used in portable atomic sensors. The static magnetic field along the measurement axis was determined from shifts in the center of the dispersive response relative to the baseline. The response distortion caused by off-axis magnetic-field components was assessed from the absorptive admixture. The equivalent additional magnetic noise was quantified from the quadrature difference in the magnetic-field amplitude spectral density (ASD) relative to the baseline. At a cell temperature of $140~^{\circ}\mathrm{C}$, the platform operated in the near-SERF regime with a $-3$~dB bandwidth of 71.5~Hz and a median ASD of 57.6~fT/$\sqrt{\mathrm{Hz}}$ over 20--70~Hz. A photodetector (PD) board and a PT1000 resistance temperature detector (RTD), both used in a portable all-optical atomic magnetometer, were evaluated using the platform. For the unpowered PD board, an orientation dependence of the static magnetic field was observed, and the degradation of the dispersive response was analyzed in terms of magnetic-field inhomogeneity. The magnetic-noise contribution was further estimated from the equivalent additional magnetic noise. In contrast, operation of the RTD readout produced only a small static magnetic field, with no measurable response degradation or additional magnetic noise. The platform can therefore guide component selection, circuit-board design, and component placement in the development of future portable atomic sensors.

\section{Zero-field optically pumped magnetometry platform}

\subsection{Configuration}

Figure~\ref{fig:setup}(a) shows a schematic of the zero-field optically pumped magnetometry platform. The atomic vapor cell, which has a cubic geometry with an inner dimension of 5.5~mm and contains $^{87}$Rb and 250~Torr of N$_2$ buffer gas,\cite{Yim2022experimental} is positioned at the center of a four-layer mu-metal shield (Twinleaf, MS-1). The cell is heated using four double-layered polyimide film heaters designed to minimize magnetic fields generated by the heating currents,\cite{Yim2018note} and its temperature is monitored using a PT1000 RTD attached to the bottom surface of the cell. Three-axis magnetic-field coils, installed inside the shield, are used to compensate for residual magnetic fields and establish near-zero-field conditions within the cell. The coil currents are supplied by current drivers (CD3-25) controlled using the analog outputs of data acquisition (DAQ) devices (National Instruments, USB-6341 and cDAQ-9185/NI-9262).

The optical system employs two distributed Bragg reflector (DBR) lasers (Photodigm, PH795DBR) in an orthogonal pump--probe configuration. The pump beam, propagating along the $+z$ axis and tuned near the $^{87}$Rb D$_1$ resonance, is expanded to a diameter of approximately 5~mm using a pair of lenses. The expanded beam passes through a linear polarizer (LP) followed by a quarter-wave plate (QWP) and enters the cell with circular polarization at a power of 14.2~mW. The probe beam, propagating along the $+x$ axis and detuned far from the $^{87}$Rb D$_1$ resonance, has a diameter of approximately 1.4~mm. The beam passes through an LP and enters the cell with linear polarization at a power of 1.2~mW. After exiting the cell, the beam is transmitted through a half-wave plate (HWP) and is split by a polarizing beam splitter (PBS). The Faraday rotation is measured using a balanced photodetector (BPD; Thorlabs, PDB210A/M), and its differential output is acquired through an analog input of a DAQ device (National Instruments, cDAQ-9185/NI-9202).

The electronic component under evaluation, hereafter referred to as the device under test (DUT), was mounted on the holder shown in Fig.~\ref{fig:setup}(b) and positioned at a location in the $xz$ plane that did not obstruct either the pump or the probe beam. The holder height was adjusted to reproduce the distance between the DUT and the cell in the portable atomic sensor.

\subsection{Theoretical model of the dispersive response}

The differential BPD signal is proportional to $P_x$, which is obtained by solving the following Bloch equation for the electron spin-polarization vector $\mathbf{P}=(P_x,P_y,P_z)$:\cite{Shah2009spin}
\begin{equation}
\frac{d\mathbf{P}}{dt}
=
\frac{1}{q}
\left[
\gamma_{\mathrm{e}}\mathbf{B}\times\mathbf{P}
+
R_{\mathrm{op}}
\left(s\hat{\mathbf{z}}-\mathbf{P}\right)
-
\boldsymbol{\Gamma}\cdot\mathbf{P}
\right],
\label{eq:bloch}
\end{equation}
where $q$ is the nuclear slowing-down factor, $\gamma_{\mathrm{e}}$ is the magnitude of the electron gyromagnetic ratio, $\mathbf{B}=(B_x,B_y,B_z)$ is the magnetic-field vector, $R_{\mathrm{op}}$ is the optical-pumping rate, $s$ is the normalized photon spin, and $\boldsymbol{\Gamma}$ is the spin-relaxation tensor. Assuming a fully circularly polarized pump beam and axial symmetry of spin relaxation about the pump axis, $s$ is set to 1 and $\boldsymbol{\Gamma}$ is taken as $\mathrm{diag}(\Gamma_{\perp},\Gamma_{\perp},\Gamma_{\parallel})$, where $\Gamma_{\perp}$ and $\Gamma_{\parallel}$ are the transverse and longitudinal spin-relaxation rates, respectively. Under steady-state conditions, $d\mathbf{P}/dt=0$, Eq.~\eqref{eq:bloch} yields
\begin{equation}
P_x(\mathbf{B})
=
\frac{R_{\mathrm{op}}}{\gamma_\mathrm{e}}
\frac{B_y}{B_y^2+B_{\mathrm{w}}^2}
+
\frac{R_{\mathrm{op}}B_xB_z}{\Gamma_2 B_{\mathrm{w}}}
\frac{B_{\mathrm{w}}}{B_y^2+B_{\mathrm{w}}^2}.
\label{eq:bloch_sol}
\end{equation}
Here, $\Gamma_1=R_{\mathrm{op}}+\Gamma_{\parallel}$ and $\Gamma_2=R_{\mathrm{op}}+\Gamma_{\perp}$ are the effective longitudinal and transverse relaxation rates, respectively, and $B_{\mathrm{w}}^2=\Gamma_1\Gamma_2/\gamma_{\mathrm{e}}^2+B_x^2+(\Gamma_1/\Gamma_2)B_z^2$. For fixed $B_x$ and $B_z$, the first term is odd in $B_y$ and represents an antisymmetric dispersive component, whereas the second term is even in $B_y$ and represents a symmetric absorptive component. The corresponding coefficients are
$C_{\mathrm{d}}=R_{\mathrm{op}}/\gamma_\mathrm{e}$ for the dispersive component and $C_{\mathrm{a}}=R_{\mathrm{op}}B_xB_z/(\Gamma_2B_{\mathrm{w}})$ for the absorptive component. To quantify the absorptive contribution arising from $B_x$ and $B_z$ relative to the dispersive component, the absorptive admixture $\eta$ is defined as
\begin{equation}
\eta
=
\frac{C_{\mathrm{a}}}{C_{\mathrm{d}}}
=
\frac{\gamma_{\mathrm{e}}B_xB_z}{\Gamma_2B_{\mathrm{w}}}.
\label{eq:admixture}
\end{equation}
For nonzero $B_x$ and $B_z$, $\lvert\eta\rvert$ increases with the magnitude of either component, while $\eta$ remains invariant under their simultaneous reversal. The strength of the $P_x$ response to small variations in $B_y$ near zero is characterized by the slope $S_{\mathrm{PB}}$, which is expressed as
\begin{equation}
S_{\mathrm{PB}}
=
\left.\frac{dP_x}{dB_y}\right|_{B_y=0}
=
\frac{R_{\mathrm{op}}}{\gamma_\mathrm{e}B_{\mathrm{w}}^2}.
\label{eq:slope}
\end{equation}
With the residual $B_x$ and $B_z$ components compensated to near zero, the magnetic field is approximated as $\mathbf{B}\simeq(0,B_y,0)$. Under this approximation, $C_{\mathrm{a}}$ and $\eta$ vanish, and Eq.~\eqref{eq:bloch_sol} reduces to a purely dispersive response to $B_y$ centered at $B_y=0$. The $y$ axis is therefore defined as the measurement axis. The response width $B_{\mathrm{w}}$, defined as the separation between the response center and either extremum, then takes the form
\begin{equation}
B_{\mathrm{w}}=\frac{\sqrt{\Gamma_1\Gamma_2}}{\gamma_{\mathrm{e}}}.
\label{eq:width}
\end{equation}
$B_{\mathrm{w}}$ sets the characteristic magnetic-field scale of the dispersive response, including the approximately linear region near $B_y=0$. At fixed $R_{\mathrm{op}}$, an increase in either $\Gamma_1$ or $\Gamma_2$ results in a broader response with a larger $B_{\mathrm{w}}$ and, as follows from Eq.~\eqref{eq:slope}, a smaller $S_{\mathrm{PB}}$.

\begin{figure}
    \centering
    \includegraphics[width=\columnwidth]{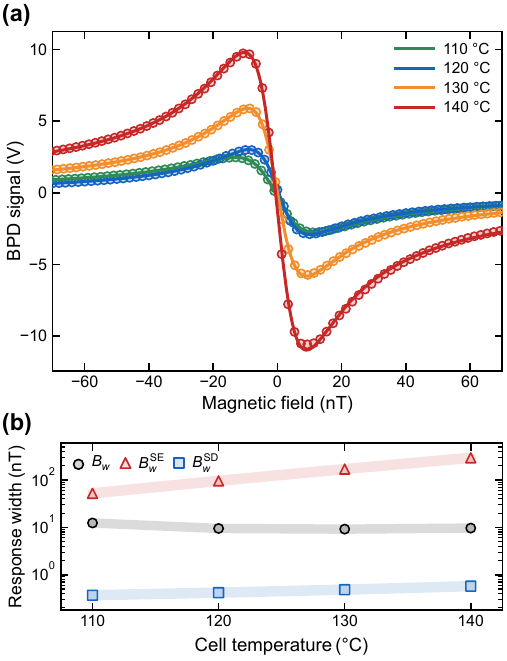}
    \caption{Temperature dependence of the dispersive response. (a) Dispersive responses measured at cell temperatures from 110 to $140~^{\circ}\mathrm{C}$. The markers represent the measured data, and the solid curves are fits to Eq.~\eqref{eq:bloch_sol}. (b) Comparison of the measured response width $B_\mathrm{w}$ with $B_\mathrm{w}^{\mathrm{SE}}$ and $B_\mathrm{w}^{\mathrm{SD}}$, the equivalent widths calculated from the spin-exchange and spin-destruction collision rates, respectively. The connecting lines highlight the temperature-dependent trends.}
    \label{fig:serf}
\end{figure}

\begin{figure*}
    \centering
    \includegraphics[width=\textwidth]{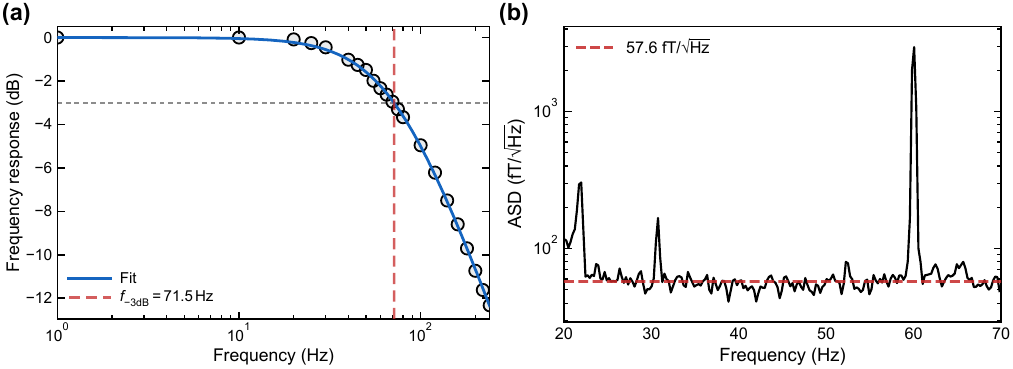}
    \caption{Performance of the zero-field optically pumped magnetometry platform. (a) Normalized frequency response measured over 1--240~Hz. The markers represent the measured data, and the solid curve is the fit to the low-pass response model. The horizontal dashed line marks the $-3$~dB level. (b) Magnetic-field amplitude spectral density (ASD) over 20--70~Hz. The dashed horizontal line marks the median ASD of 57.6~fT/$\sqrt{\mathrm{Hz}}$.}
    \label{fig:performance}
\end{figure*}

\subsection{Magnetic characterization procedure}

The magnetic interference associated with each DUT was characterized by comparing measurements acquired under baseline and DUT conditions. A baseline response was first obtained without a DUT by sweeping the applied magnetic field $B_y$ from $-70$ to $+70$~nT in 2~nT increments. The response was then fitted using Eq.~\eqref{eq:bloch_sol} expressed on a voltage scale to determine the response center $B_0$, response width $B_{\mathrm{w}}$, and response slope $S_{\mathrm{VB}}\propto S_{\mathrm{PB}}$ at $B_y=B_0$. The applied field $B_y$ was subsequently set to $B_0$, and the output voltage was recorded as a time series. The voltage ASD was calculated from the time series sampled at 1~kHz using Welch's method (Hann window, 4096-sample segments, and 50\% overlap) and then divided by $\lvert S_{\mathrm{VB}}\rvert$ to obtain the magnetic-field ASD. To reduce the influence of isolated narrowband peaks, the median of the magnetic-field ASD within the analysis frequency band was used as a representative estimate of the magnetic-noise level. After the DUT was mounted at the specified measurement position, the same procedure was repeated under otherwise identical operating conditions, with the time series recorded at $B_y=B_{0,\mathrm{DUT}}$.

The static magnetic field generated by the DUT, as measured along the measurement axis, was determined from the shift in $B_0$ relative to the baseline:
\begin{equation}
B_{\mathrm{DUT},y}
=
B_{0,\mathrm{DUT}}
-
B_{0,\mathrm{baseline}}.
\label{eq:static}
\end{equation}
The response characteristics of the platform were further assessed in terms of absorptive admixture $\eta$ and response width $B_{\mathrm{w}}$, defined in Eqs.~\eqref{eq:admixture} and \eqref{eq:width}, respectively. As described by Eq.~\eqref{eq:bloch_sol}, an absorptive component can arise when both $B_x$ and $B_z$ are nonzero, resulting in asymmetric distortion of the dispersive line shape. An increase in $B_{\mathrm{w}}$ accompanied by a reduction in $\lvert S_{VB}\rvert$ relative to the baseline was interpreted as degradation of the dispersive response consistent with increased magnetic-field inhomogeneity across the atomic vapor cell.\cite{Li2024comprehensive}

The equivalent additional magnetic noise $\mathcal{N}_{\mathrm{add}}$ was determined from the quadrature difference between the median magnetic-field ASDs measured under the DUT and baseline conditions, with the noise contributions assumed to be mutually uncorrelated:
\begin{equation}
\begin{split}
\mathcal{N}_{\mathrm{add}}^{2}
&=
\mathcal{N}_{\mathrm{DUT}}^{2}
-
\mathcal{N}_{\mathrm{baseline}}^{2}
\\
&=
K\left(
\mathcal{N}_{V,\mathrm{PSN}}^{2}
+
\mathcal{N}_{V,\mathrm{dark}}^{2}
\right)
+
\mathcal{N}_{B,\mathrm{DUT}}^{2}.
\end{split}
\label{eq:noise}
\end{equation}
The baseline median ASD $\mathcal{N}_{\mathrm{baseline}}$ comprises nonmagnetic system noise, represented in the voltage domain by photon shot noise $\mathcal{N}_{V,\mathrm{PSN}}$ and dark electronic noise $\mathcal{N}_{V,\mathrm{dark}}$, together with background magnetic noise $\mathcal{N}_{B,\mathrm{bg}}$, and is expressed as $\mathcal{N}_{\mathrm{baseline}}^{2}=(\mathcal{N}_{V,\mathrm{PSN}}^{2}+\mathcal{N}_{V,\mathrm{dark}}^{2})/\lvert S_{\mathrm{VB},\mathrm{baseline}}\rvert^{2}+\mathcal{N}_{B,\mathrm{bg}}^{2}$. Under the DUT condition, the median ASD $\mathcal{N}_{\mathrm{DUT}}$ additionally includes a DUT-associated magnetic-noise contribution $\mathcal{N}_{B,\mathrm{DUT}}$ and therefore takes the form $\mathcal{N}_{\mathrm{DUT}}^{2}=(\mathcal{N}_{V,\mathrm{PSN}}^{2}+\mathcal{N}_{V,\mathrm{dark}}^{2})/\lvert S_{\mathrm{VB},\mathrm{DUT}}\rvert^{2}+\mathcal{N}_{B,\mathrm{bg}}^{2}+\mathcal{N}_{B,\mathrm{DUT}}^{2}$. Because the nonmagnetic system noise is common to both conditions, its contribution to the magnetic-field ASD varies with the response slope $\lvert S_{\mathrm{VB}}\rvert$, and the resulting difference is accounted for by the coefficient $K=1/\lvert S_{\mathrm{VB},\mathrm{DUT}}\rvert^{2}-1/\lvert S_{\mathrm{VB},\mathrm{baseline}}\rvert^{2}$. In contrast, the background magnetic noise is common to both conditions and therefore cancels in the quadrature difference. Accordingly, the magnetic-field noise associated with the DUT can be estimated from Eq.~\eqref{eq:noise} as $\mathcal{N}_{B,\mathrm{DUT}}=\left[\mathcal{N}_{\mathrm{add}}^{2}-K\left(\mathcal{N}_{V,\mathrm{PSN}}^{2}+\mathcal{N}_{V,\mathrm{dark}}^{2}\right)\right]^{1/2}$.

\subsection{Operating temperature}

Figure~\ref{fig:serf}(a) shows the dispersive responses measured at four atomic vapor cell temperatures ranging from 110 to 140~$^{\circ}\mathrm{C}$. Measurements were limited to 140~$^{\circ}\mathrm{C}$ to avoid saturation of the BPD and remain within the allowable operating range of the heater. $\lvert S_{VB}\rvert$ increased monotonically with temperature, rising from 0.420~V/nT at 110~$^{\circ}\mathrm{C}$ to 0.622~V/nT at 120~$^{\circ}\mathrm{C}$, 1.27~V/nT at 130~$^{\circ}\mathrm{C}$, and 2.15~V/nT at 140~$^{\circ}\mathrm{C}$. Meanwhile, $B_\mathrm{w}$ decreased from 12.4~nT at 110~$^{\circ}\mathrm{C}$ to 9.49 and 9.19~nT at 120 and 130~$^{\circ}\mathrm{C}$, respectively, before slightly increasing to 9.66~nT at 140~$^{\circ}\mathrm{C}$.

To assess whether these temperature-dependent responses were consistent with SERF operation, the measured $B_\mathrm{w}$ values were compared with the spin-destruction-equivalent width $B_\mathrm{w}^{\mathrm{SD}}$ and the spin-exchange-equivalent width $B_\mathrm{w}^{\mathrm{SE}}$, as shown in Fig.~\ref{fig:serf}(b). Both equivalent widths were obtained by dividing the corresponding collision rates by $\gamma_e=2\pi\times28~\mathrm{Hz/nT}$.\cite{Shah2009spin} The spin-destruction rate was calculated as $R_\mathrm{SD}=\sum_{i=\mathrm{Rb,N_2}}n_i\sigma_\mathrm{SD}^{\mathrm{Rb}-i}\bar{v}_\mathrm{rel}^{\mathrm{Rb}-i}$,\cite{Liu2017polarization} where $n_i$ is the number density of collision partner $i$,\cite{Kim2025aging} $\sigma_\mathrm{SD}^{\mathrm{Rb}-i}$ is the Rb--$i$ spin-destruction cross section,\cite{Chen2007spin} and $\bar{v}_\mathrm{rel}^{\mathrm{Rb}-i}$ is the corresponding mean relative speed.\cite{Gentile2017optically} The resulting $B_\mathrm{w}^{\mathrm{SD}}$ increased monotonically from 0.37~nT at 110~$^{\circ}\mathrm{C}$ to 0.57~nT at 140~$^{\circ}\mathrm{C}$, representing an increase by a factor of 1.6. At all temperatures, $B_\mathrm{w}^{\mathrm{SD}}$ remained more than an order of magnitude smaller than the measured $B_\mathrm{w}$, indicating that the measured response width includes substantial contributions from relaxation and broadening mechanisms beyond spin-destruction collisions. The spin-exchange collision rate was calculated as $R_\mathrm{SE}=n_\mathrm{Rb}\sigma_\mathrm{SE}^{\mathrm{Rb}-\mathrm{Rb}}\bar{v}_\mathrm{rel}^{\mathrm{Rb}-\mathrm{Rb}}$,\cite{Walker1997spin} where $\sigma_\mathrm{SE}^{\mathrm{Rb}-\mathrm{Rb}}$ is the Rb--Rb spin-exchange cross section.\cite{Gibbs1967spin} The resulting $B_\mathrm{w}^{\mathrm{SE}}$ increased monotonically from 52.5~nT at 110~$^{\circ}\mathrm{C}$ to 294.5~nT at 140~$^{\circ}\mathrm{C}$, representing an increase by a factor of 5.6.

The measured $B_\mathrm{w}$ decreased and then remained nearly constant, whereas $B_\mathrm{w}^{\mathrm{SE}}$, calculated assuming unsuppressed spin-exchange relaxation, increased with temperature and was substantially larger than the measured $B_\mathrm{w}$ at all temperatures. An upper-bound estimate of the Larmor angular frequency within the approximately linear response region was obtained as $\omega_L(B_\mathrm{w})=(\gamma_e/q)B_\mathrm{w}$, with $q\approx4$.\cite{Korver2013suppression} The ratio of $R_\mathrm{SE}$ to $\omega_L(B_\mathrm{w})$ increased monotonically with temperature, from approximately 16.9 at 110~$^{\circ}\mathrm{C}$ to 121.8 at 140~$^{\circ}\mathrm{C}$. At 140~$^{\circ}\mathrm{C}$, $\omega_L(B_\mathrm{w})$ was much smaller than $R_\mathrm{SE}$, while no corresponding response broadening was observed, consistent with suppression of spin-exchange relaxation in the near-SERF regime. This temperature was selected for subsequent measurements because it yielded the largest $\lvert S_{VB}\rvert$ while $B_\mathrm{w}$ remained near its minimum.

\subsection{Performance}

The frequency response of the platform was measured to determine its bandwidth, as shown in Fig.~\ref{fig:performance}(a). A sinusoidal magnetic field was applied using the $y$-axis coil, with its amplitude fixed at 0.33~nT to ensure operation within the linear response region. The corresponding output-voltage amplitude was measured as the field frequency was varied. The normalized response exhibited low-pass behavior and was fitted to $R(f)=[1+(f/f_{-3\mathrm{dB}})^{2n}]^{-1/2}$,\cite{Ma2024ultrasensitive} where $f_{-3\mathrm{dB}}$ is the $-3$~dB cutoff frequency and $n$ characterizes the roll-off. The fit yielded $f_{\mathrm{-3dB}}=71.5$~Hz and $n=1.15$, indicating an approximately first-order low-pass response. Accordingly, the analysis frequency band for the subsequent ASD evaluation was set to 20--70~Hz, with the lower limit chosen to exclude low-frequency technical noise and the upper limit set just below the measured cutoff frequency.

The magnetic-field sensitivity of the platform was evaluated in terms of the ASD, as shown in Fig.~\ref{fig:performance}(b). The median ASD over 20--70~Hz was 57.6~fT/$\sqrt{\mathrm{Hz}}$, with a minimum value of 41.0~fT/$\sqrt{\mathrm{Hz}}$ at 42~Hz. The pronounced peak at 60~Hz is attributed to power-line interference. The response parameters and ASD reported here characterize the baseline performance of the platform under the optimized operating conditions.

\begin{figure}
    \centering
    \includegraphics[width=\columnwidth]{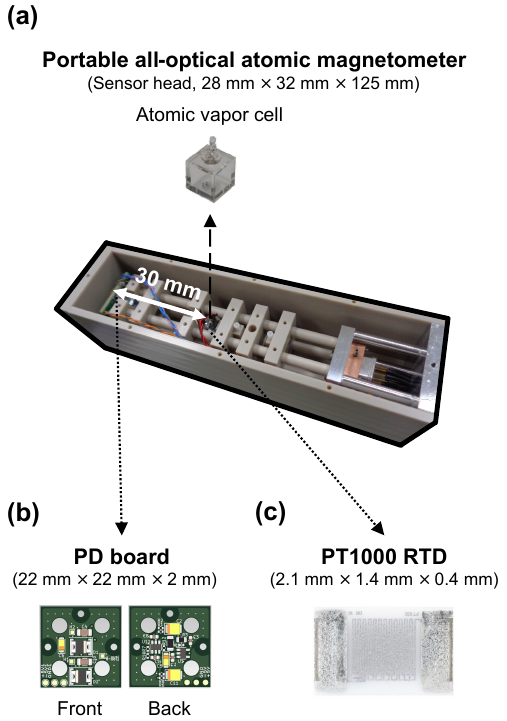}
    \caption{Configuration of the portable all-optical atomic magnetometer. (a) Photograph showing the interior of the sensor head. The photodetector (PD) board is positioned approximately 30~mm from the cell, whereas the PT1000 resistance temperature detector (RTD) is attached directly beneath the cell to monitor its temperature. (b) Front and back sides of the PD board. (c) Photograph of the PT1000 RTD.}
    \label{fig:sensor}
\end{figure}

\section{Application to a portable atomic magnetometer}

The portable all-optical atomic magnetometer, shown in Fig.~\ref{fig:sensor}(a), uses a single laser beam for pulsed operation and measures the scalar magnetic field from the FID signal.\cite{Yoon2024laser} The sensor head has dimensions of $28~\mathrm{mm}\times32~\mathrm{mm}\times125~\mathrm{mm}$ and achieved a sensitivity of 1.98~pT/$\sqrt{\mathrm{Hz}}$ over 5--100~Hz under magnetically shielded conditions.\cite{Kim2026magnetic} Based on our previous studies, the atomic vapor cell heater was designed to reduce magnetic-field generation,\cite{Yim2018note} and a laser with low-magnetic packaging was used.\cite{Kim2023low} Accordingly, the static magnetic fields and magnetic noise of the two remaining electronic components, the PD board and PT1000 RTD, were characterized at their respective distances from the cell.

\begin{figure*}
    \centering
    \includegraphics[width=\textwidth]{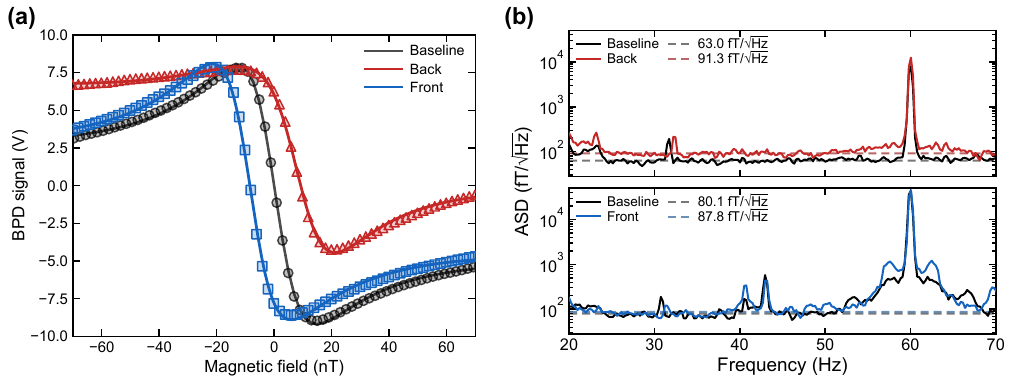}
    \caption{Magnetic characterization of the unpowered PD board at a board-to-cell distance of 30~mm in the back- and front-facing orientations. The baseline refers to measurements performed without the PD board. (a) Dispersive responses for the baseline and the two board orientations. The markers represent the measured data, and the solid curves are fits to Eq.~(\ref{eq:bloch_sol}). For clarity, the two baseline data sets were averaged point by point for display, while the individual data sets were used for quantitative analysis. (b) Magnetic-field ASDs for the back-facing (upper plot) and front-facing (lower plot) orientations and their respective baselines. The dashed horizontal lines indicate the median ASDs over 20--70~Hz.}
    \label{fig:board}
\end{figure*}

\subsection{Photodetector board}

The PD board, shown in Fig.~\ref{fig:sensor}(b), has dimensions of $22~\mathrm{mm}\times22~\mathrm{mm}\times2~\mathrm{mm}$ and different circuit layouts on its front and back sides. The front side, which faces the atomic vapor cell in the sensor head, contains two photodiodes and three capacitors, whereas the back side comprises two operational amplifiers, nine resistors, and eight capacitors. This difference in the distribution of conductive traces and magnetic materials within the circuit elements, such as the nickel barrier layers in the capacitor terminations, may produce different magnetic fields at the atomic vapor cell depending on the board orientation. Measurements were therefore performed with the unpowered board in two orientations, with either the front or back side facing the atomic vapor cell. The two cases are referred to as the front-facing and back-facing orientations, respectively. The board was mounted on a nonmagnetic holder at a board-to-cell distance of 30~mm, corresponding to its placement in the portable all-optical atomic magnetometer.

\subsubsection{Results}

Figure~\ref{fig:board}(a) presents the dispersive responses measured in the back-facing and front-facing orientations. In the back-facing orientation, the static magnetic field generated by the PD board was 8.7~nT, as determined using Eq.~\eqref{eq:static}. $B_\mathrm{w}$ increased from 13.0 to 15.3~nT, whereas $\lvert S_\mathrm{VB}\rvert$ decreased from 1.3 to 0.8~V/nT. The dispersive line shape was also distorted, with pronounced flattening of the left lobe. In the front-facing orientation, by contrast, the static magnetic field generated by the PD board was $-7.9$~nT. $B_\mathrm{w}$ increased from 13.4 to 13.7~nT, while $\lvert S_\mathrm{VB}\rvert$ remained unchanged at 1.3~V/nT.

Figure~\ref{fig:board}(b) compares the magnetic-field ASDs obtained in the back-facing and front-facing orientations with their respective baselines. Each baseline was measured without the PD board. Because the baseline ASD increased over the course of the sequential measurements, each board orientation was evaluated relative to its separately acquired baseline. In the back-facing orientation, the median ASD over 20--70~Hz increased from 63.0 to 91.3~fT/$\sqrt{\mathrm{Hz}}$ [Fig.~\ref{fig:board}(b), upper plot]. The equivalent additional magnetic noise, calculated using Eq.~\eqref{eq:noise}, was 66.1~fT/$\sqrt{\mathrm{Hz}}$. In the front-facing orientation, the median ASD increased from 80.1 to 87.8~fT/$\sqrt{\mathrm{Hz}}$, corresponding to an equivalent additional magnetic noise of 36.0~fT/$\sqrt{\mathrm{Hz}}$ [Fig.~\ref{fig:board}(b), lower plot].

\subsubsection{Discussion}

\begin{figure}
    \centering
    \includegraphics[width=\columnwidth]{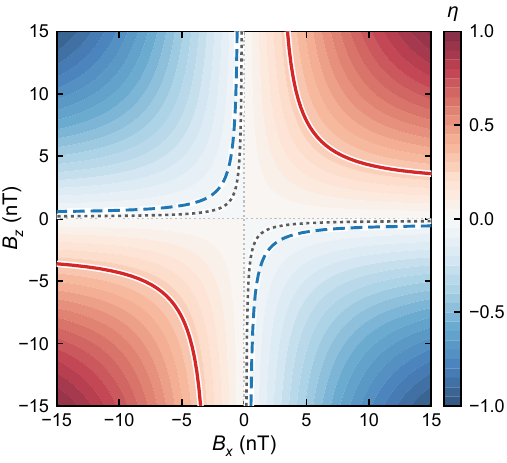}
    \caption{Distribution of the absorptive admixture as a function of $B_x$ and $B_z$, with gray dotted (baseline, $-0.016$), red solid (back-facing, $0.29$), and blue dashed (front-facing, $-0.047$) contours. The color map was calculated using Eq.~\eqref{eq:admixture} with $\Gamma_1=1.6\times10^3~\mathrm{s}^{-1}$ and $\Gamma_2=1.8\times10^3~\mathrm{s}^{-1}$.}
    \label{fig:board_eta}
\end{figure}

The static magnetic field from the PD board had a $y$-component of comparable magnitude but opposite sign in the two orientations. This sign reversal is consistent with a net magnetic moment associated with the board, whose orientation relative to the cell is reversed when the board is flipped. Despite the comparable $y$-component magnitudes, the dispersive response was substantially more degraded in the back-facing orientation, exhibiting pronounced asymmetric distortion. As shown by Eq.~\eqref{eq:bloch_sol}, nonzero $B_x$ and $B_z$ give rise to a symmetric absorptive component, whose contribution relative to the dispersive component is quantified by $\eta$ in Eq.~\eqref{eq:admixture}. Figure~\ref{fig:board_eta} shows the distribution of $\eta$ as a function of $B_x$ and $B_z$, with contours corresponding to the baseline, back-facing, and front-facing conditions. The $\eta$ distribution was calculated using $\Gamma_2\simeq2\pi qf_{-3\mathrm{dB}}=1.8\times10^3~\mathrm{s^{-1}}$ under the first-order spin-response approximation, estimated from $f_{-3\mathrm{dB}}$ in Fig.~\ref{fig:performance}(a), and $\Gamma_1=1.6\times10^3~\mathrm{s^{-1}}$, obtained from Eq.~\eqref{eq:width} using $B_\mathrm{w}$ at $140\,^{\circ}\mathrm{C}$ in Fig.~\ref{fig:serf}(b). The distribution is centrosymmetric about the origin ($B_x=B_z=0$), where $\eta=0$ and the response is purely dispersive. Each contour represents the possible combinations of $B_x$ and $B_z$ consistent with the measured $\eta$, rather than unique values of the two field components. In the back-facing orientation, $\eta$ changed from the baseline value of $-0.016$ to $0.29$. Compared with the baseline contour, the back-facing contour lies farther from both axes, reflecting a larger absorptive admixture consistent with the pronounced response distortion. The sign reversal indicates that either $B_x$ or $B_z$ changed sign, although $\eta$ alone does not identify which component. In the front-facing orientation, $\eta$ shifted modestly to $-0.047$, consistent with the largely preserved dispersive response.

The increase in $B_{\mathrm{w}}$ was observed in both orientations but was more pronounced in the back-facing orientation. This broadening likely arose from magnetic-field inhomogeneity across the atomic vapor cell associated with the spatial arrangement of magnetic materials within the board. The simplified model presented in Appendix~\ref{app:gradient}, given by Eq.~\eqref{eq:gradient_rel}, shows that the quadrature difference in $B_\mathrm{w}$ between each board orientation and its baseline can be used as an indicator for comparing magnetic-field-gradient strengths between the two orientations. The larger quadrature difference for the back-facing orientation indicates greater magnetic-field inhomogeneity across the atomic vapor cell than for the front-facing orientation.

To estimate the contribution of magnetic-field noise associated with the PD board to the increase in the median ASD, the photon shot noise and dark electronic noise were evaluated in magnetic-field units. The ASD of photon shot noise is given by $\sqrt{2eg^2RP}/\lvert S_{\mathrm{VB}}\rvert$, where $e$ is the elementary charge, $g=5\times10^5~\mathrm{V/A}$ is the transimpedance gain for a high-impedance load, $R=0.55~\mathrm{A/W}$ is the photodiode responsivity at 795~nm, and $P$ is the total optical power incident on the two photodiodes. For the back-facing orientation, the photon shot noise increased from 5.6~fT/$\sqrt{\mathrm{Hz}}$ to 9.1~fT/$\sqrt{\mathrm{Hz}}$, reflecting the decrease in $\lvert S_{\mathrm{VB}}\rvert$ accompanying the response degradation. The dark electronic noise, determined from the ASD measured with no incident light, likewise increased from 14.0~fT/$\sqrt{\mathrm{Hz}}$ to 22.7~fT/$\sqrt{\mathrm{Hz}}$. The corresponding quadrature differences, obtained following Eq.~\eqref{eq:noise}, were 7.2~fT/$\sqrt{\mathrm{Hz}}$ for photon shot noise and 17.9~fT/$\sqrt{\mathrm{Hz}}$ for dark electronic noise. By subtracting the photon shot noise and dark electronic noise contributions in quadrature from the equivalent additional magnetic noise of 66.1~fT/$\sqrt{\mathrm{Hz}}$, the magnetic-noise contribution associated with the PD board was estimated to be 63.2~fT/$\sqrt{\mathrm{Hz}}$ in this orientation. For the front-facing orientation, the photon shot noise and dark electronic noise remained at 5.6 and 14.0~fT/$\sqrt{\mathrm{Hz}}$, respectively, consistent with the unchanged $\lvert S_{\mathrm{VB}}\rvert$. Accordingly, the magnetic-noise contribution associated with the PD board was estimated to be 36.0~fT/$\sqrt{\mathrm{Hz}}$, equal to the equivalent additional magnetic noise.

A possible source of this contribution is thermal magnetic-field noise arising from Johnson-current fluctuations in the conductive traces and magnetization fluctuations in magnetic materials within the PD board. Based on the thin-disk model,\cite{Lee2008calculation} with each of the two patterned copper layers approximated as a nonmagnetic conducting disk and white-noise behavior assumed at low frequencies, the Johnson-current magnetic noise from each layer was estimated as $\delta B_{\mathrm{J}}=(1/\sqrt{8\pi})[\mu_0\sqrt{k_{\mathrm{B}}T\sigma t}/a]/(1+a^2/r^2)$. Here, $\mu_0$ is the vacuum permeability, $k_{\mathrm{B}}$ is the Boltzmann constant, $T=413$~K is the disk temperature set equal to the cell operating temperature, $\sigma=5.8\times10^7~\mathrm{S/m}$ is the electrical conductivity of copper, $t=17.5~\mu\mathrm{m}$ is the thickness of each copper layer, $a=30$~mm is the axial distance from the disk to the vapor cell, and $r=12.4$~mm is the equivalent disk radius for the board. The estimated Johnson-current magnetic noise from each copper layer was 2.9~fT/$\sqrt{\mathrm{Hz}}$. Assuming uncorrelated noise from the two layers, their contributions were added in quadrature, yielding a total Johnson-current magnetic noise of 4.1~fT/$\sqrt{\mathrm{Hz}}$. This value is markedly smaller than the magnetic-noise contributions of 63.2 and 36.0~fT/$\sqrt{\mathrm{Hz}}$, indicating that Johnson-current fluctuations in the conductive traces alone cannot account for the magnetic noise, while thermal magnetization fluctuations in magnetic materials within the PD board may provide an additional contribution. Under geomagnetic-field conditions, the magnetic-noise contribution associated with the PD board could be larger than that estimated under the near-zero-field conditions considered here owing to additional fluctuations in field-induced magnetization.

\begin{figure*}
    \centering
    \includegraphics[width=\textwidth]{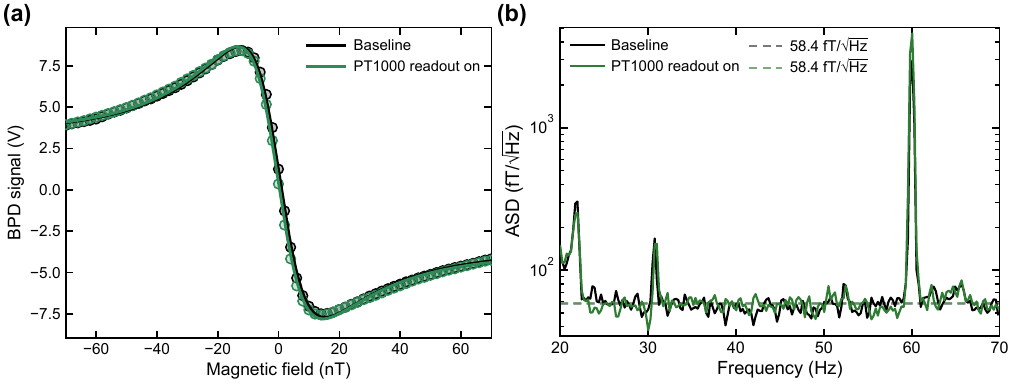}
    \caption{Magnetic characterization under different PT1000 RTD readout conditions. (a) Dispersive responses measured under the baseline and readout-enabled conditions. The baseline is the condition with the RTD readout disabled. The markers represent the measured data, and the solid curves are fits to Eq.~(\ref{eq:bloch_sol}). (b) Corresponding magnetic-field ASDs. The dashed horizontal lines indicate the median ASD values over 20--70~Hz.}
    \label{fig:RTD}
\end{figure*}

\subsection{Resistance temperature detector}

The PT1000 RTD, shown in Fig.~\ref{fig:sensor}(c), has dimensions of $2.1~\mathrm{mm}\times1.4~\mathrm{mm}\times0.4~\mathrm{mm}$ and comprises a platinum resistive element arranged in a meander pattern on a ceramic substrate.\cite{YageoNexensosSMD0805FC} The RTD was attached directly beneath the atomic vapor cell for continuous temperature monitoring, as in the portable all-optical atomic magnetometer. The readout-disabled condition was used as the baseline and compared with the readout-enabled condition to assess the magnetic interference associated with the readout current.

Figure~\ref{fig:RTD}(a) presents the dispersive responses measured under the baseline and readout-enabled conditions. The static magnetic field generated by the readout current was $-0.6$~nT, as determined using Eq.~\eqref{eq:static}. Meanwhile, $B_\mathrm{w}$ and $\lvert S_{VB}\rvert$ remained nearly unchanged at 13.6~nT and 1.2~V/nT, respectively, indicating no measurable response degradation due to additional magnetic-field inhomogeneity across the cell. $\eta$ also changed only slightly, from $-0.017$ under the baseline condition to $-0.018$ with the RTD readout enabled. Figure~\ref{fig:RTD}(b) compares the corresponding magnetic-field ASDs. The median ASD over 20--70~Hz was 58.4~fT/$\sqrt{\mathrm{Hz}}$ under both conditions, indicating no measurable increase in magnetic noise. This negligible magnetic interference is consistent with the low readout current (0.1--0.3~mA) and partial cancellation between the magnetic fields generated by oppositely directed currents in adjacent segments of the meander pattern. For direct comparison, the magnetic characterization results for the PD board and PT1000 RTD are summarized in Appendix~\ref{app:results}.

\section{Conclusion}
We developed a zero-field optically pumped magnetometry platform and applied it to the magnetic characterization of a PD board and an RTD used in a portable atomic magnetometer. The static magnetic-field component along the measurement axis was determined from shifts in the center of the dispersive response relative to the baseline, while response distortion caused by off-axis magnetic-field components was assessed from the absorptive admixture. The equivalent additional magnetic noise was quantified from the quadrature difference in the ASD. At a cell temperature of $140~^{\circ}\mathrm{C}$, the platform operated in the near-SERF regime and exhibited a $-3$~dB bandwidth of 71.5~Hz and a median ASD of 57.6~fT/$\sqrt{\mathrm{Hz}}$ over 20--70~Hz. Measurements of the PD board revealed an orientation dependence in the static magnetic field, with shifts of $-7.9$~nT in the front-facing orientation and 8.7~nT in the back-facing orientation. Additionally, the back-facing orientation exhibited pronounced response degradation consistent with magnetic-field inhomogeneity across the atomic vapor cell. The magnetic-noise contributions associated with the PD board were estimated at 36.0 and 63.2~fT/$\sqrt{\mathrm{Hz}}$ in the front- and back-facing orientations, respectively, accounting for most of the ASD increase in both orientations. In contrast, operation of the PT1000 RTD readout generated only a static magnetic field of $-0.6$~nT, with no measurable degradation of the dispersive response or increase in the ASD. By characterizing the static magnetic fields and magnetic noise associated with nearby electronic components, the platform can guide component selection, circuit-board design, and component placement for future portable atomic magnetometers targeting sensitivities below 0.1~pT/$\sqrt{\mathrm{Hz}}$.

\begin{acknowledgments}
This work was supported by a grant from the Agency for Defense Development, funded by the Government of the Republic of Korea (No. 915098102).
\end{acknowledgments}

\section*{AUTHOR DECLARATIONS}
\subsection*{Conflict of Interest}
The authors have no conflicts to disclose.

\subsection*{Author Contributions}
\noindent\textbf{Hyeonjae Kim}: Data curation (lead); Formal analysis (equal); Investigation (lead); Methodology (equal); Software (lead); Validation (equal); Visualization (lead); Writing -- original draft (lead); Writing -- review and editing (equal).
\textbf{Sangkyung Lee}: Conceptualization (equal); Formal analysis (equal); Investigation (supporting); Methodology (equal); Resources (equal); Supervision (lead); Validation (equal); Writing -- review and editing (equal).
\textbf{Sin Hyuk Yim}: Conceptualization (equal); Resources (equal); Writing -- review and editing (supporting).
\textbf{Taek Jeong}: Resources (supporting); Visualization (supporting); Writing -- review and editing (supporting).
\textbf{Younghoon Lim}: Resources (supporting); Writing -- review and editing (supporting).

\section*{Data Availability}
The data that support the findings of this study are available from the corresponding author upon reasonable request.

\appendix
\section{Response broadening induced by magnetic-field gradients}
\label{app:gradient}

Consider near-zero-field conditions in which $B_y$ has a dominant one-dimensional first-order gradient $G_j=\partial B_y/\partial u$, where $u$ is the coordinate along the gradient direction and $j$ denotes the baseline or DUT condition. With the atomic vapor cell approximated as a sphere of radius $r_{\mathrm{c}}$, the longitudinal and transverse gradient-relaxation rates derived by Fang et al.\cite{Fang2022analysis} can be incorporated into Eq.~\eqref{eq:width}. The gradient-induced relaxation rates are taken to be small compared with $\Gamma_1$ and $\Gamma_2$. The resulting term proportional to $G_j^4$ is therefore neglected, giving
\begin{subequations}
\begin{equation}
B_{\mathrm{w},j}^2
=
\frac{
\left(\Gamma_1+\Gamma_{1\Delta B,j}\right)
\left(\Gamma_2+\Gamma_{2\Delta B,j}\right)
}{
\gamma_e^2
}
\simeq
B_{\mathrm{w}}^2+\kappa_G G_j^2,
\end{equation}
\begin{equation}
\kappa_G
=
\frac{
r_{\mathrm{c}}^4q\left(\Gamma_1+2\Gamma_2\right)
}{D}
\left[
\sum_n
\frac{1}{
x_{1n}^4\left(x_{1n}^2-2\right)
}
\right].
\end{equation}
\end{subequations}
Here, $\Gamma_{1\Delta B,j}$ and $\Gamma_{2\Delta B,j}$ denote the longitudinal and transverse relaxation rates induced by the magnetic-field gradient under condition $j$, respectively. $\kappa_G$ is the corresponding gradient-broadening coefficient. $D$ is the diffusion coefficient of Rb atoms in the buffer gas, and $x_{1n}$ denotes the dimensionless spatial eigenvalue for diffusion mode $n$ under the spherical-cell diffusion boundary conditions. Subtracting the baseline relation from the DUT relation and rearranging results in
\begin{equation}
\sqrt{
G_{\mathrm{DUT}}^2-G_{\mathrm{baseline}}^2
}
\simeq
\frac{1}{\sqrt{\kappa_G}}
\sqrt{
B_{\mathrm{w},\mathrm{DUT}}^2-B_{\mathrm{w},\mathrm{baseline}}^2
}.
\label{eq:gradient_rel}
\end{equation}
If the baseline gradient is negligible ($G_{\mathrm{baseline}}\simeq0$), the left-hand side reduces to $\lvert G_{\mathrm{DUT}}\rvert$. With $\kappa_G$ accurately determined using independently characterized diffusion and relaxation parameters within a refined model that accounts for the actual cell geometry, the gradient magnitude under the DUT condition could be quantitatively estimated from the measured broadening of the dispersive response. Here, $\sqrt{B_{\mathrm{w},\mathrm{DUT}}^2-B_{\mathrm{w},\mathrm{baseline}}^2}$ is instead used as a relative indicator of gradient strength among DUT configurations, based on its proportionality to the DUT-associated gradient contribution.

\section{Magnetic characterization results}
\label{app:results}

The magnetic characterization results for the PD board and PT1000 RTD used in the portable all-optical atomic magnetometer are summarized in Table~\ref{tab:results}.

\begin{table}[H]
\caption{Summary of the magnetic characterization results for the unpowered PD board and PT1000 RTD with its readout enabled. The response widths under the baseline and DUT conditions are listed seperately. The dash in the $N_{B,\mathrm{DUT}}$ entry for the PT1000 RTD indicates that no measurable magnetic-noise contribution was observed.}
\label{tab:results}
\centering
\begingroup
\setlength{\tabcolsep}{2.5pt}
\renewcommand{\arraystretch}{1.15}
\setcellgapes{2.5pt}
\makegapedcells
\begin{ruledtabular}
\begin{tabular}{cccccc}
\makecell[c]{Configuration}
&
\makecell[c]{$B_{\mathrm{DUT},y}$\\(nT)}
&
\makecell[c]{$\eta_{\mathrm{DUT}}$}
&
\makecell[c]{$B_{\mathrm{w},\mathrm{baseline}}$\\(nT)}
&
\makecell[c]{$B_{\mathrm{w},\mathrm{DUT}}$\\(nT)}
&
\makecell[c]{$N_{B,\mathrm{DUT}}$\\(fT/$\sqrt{\mathrm{Hz}}$)}
\\
\hline
\noalign{\vskip 2.5pt}
\makecell[c]{PD board\\Back-facing}
&
\makecell[c]{8.7}
&
\makecell[c]{0.29}
&
\makecell[c]{13.0}
&
\makecell[c]{15.3}
&
\makecell[c]{63.2}
\\
\makecell[c]{PD board\\Front-facing}
&
\makecell[c]{$-7.9$}
&
\makecell[c]{$-0.047$}
&
\makecell[c]{13.4}
&
\makecell[c]{13.7}
&
\makecell[c]{36.0}
\\
\makecell[c]{PT1000 RTD\\Readout enabled}
&
\makecell[c]{$-0.6$}
&
\makecell[c]{$-0.018$}
&
\makecell[c]{13.6}
&
\makecell[c]{13.6}
&
\makecell[c]{---}
\\
\end{tabular}
\end{ruledtabular}
\endgroup
\end{table}

\section*{References}
\bibliography{aipsamp}

\end{document}